\documentclass[fleqn,usenatbib]{mnras}

\usepackage[T1]{fontenc}

\DeclareRobustCommand{\VAN}[3]{#2}
\let\VANthebibliography\thebibliography
\def\thebibliography{\DeclareRobustCommand{\VAN}[3]{##3}\VANthebibliography}

\usepackage{graphicx}	% Including figure files
\usepackage{amsmath}	% Advanced maths commands
\usepackage{amssymb}	% Extra maths symbols

\title[Minimal Timescales]{A Minimal Timescale for the Continuum in 4U 1608-52 and Aql X-1}

\author[K. Mohamed et al.]{
K. Mohamed,$^{1,3}$
E. Sonbas,$^{2,3}$\thanks{E-mail: edasonbas@gmail.com}
K. S. Dhuga,$^{3}$
E. G\"o\u{g}\"u\c{s},$^{4}$
A. Tuncer,$^{5}$
N. N. Abd Allah,$^{1}$
\newauthor and A. Ibrahim$^{6}$ 
\\
$^{1}$Department of Physics, Sohag University, Egypt \\
$^{2}$Adiyaman University, Department of Physics, 02040 Adiyaman, Turkey\\
$^{3}$Department of Physics, The George Washington University, Washington, DC 20052, USA\\
$^{4}$Sabanc\i~University, Orhanl\i~- Tuzla, Istanbul 34956, Turkey\\
$^{5}$Istanbul University Science Faculty, Department of Astronomy and Space Sciences, 34119, University-Istanbul, Turkey\\
$^{6}$Department of Physics, Faculty of Science, Cairo University, Egypt
}

\date{Accepted XXX. Received YYY; in original form ZZZ}

\pubyear{2020}

\begin{document}
\label{firstpage}
\pagerange{\pageref{firstpage}--\pageref{lastpage}}
\maketitle

\begin{abstract}
\noindent
Similar to black-hole X-ray binary (BHXRB) transients, hysteresis-like state transitions are also seen in some neutron-star X-ray binaries (NSXRBs). Using a method based on wavelets and lightcurves constructed from archival RXTE observations, we extract a minimal timescale over the complete range of transitions for 4U 1608-52 during the 2002 and 2007 outbursts and the 1999 and 2000 outbursts for Aql X-1. We present evidence for a strong positive correlation between this minimal timescale and a similar timescale extracted from the corresponding power spectra of these sources. 
\end{abstract}


%%%%%%%%%%%%%%%%%%%%%%%%%%%%%%%%%%%%%%%%%%%%%%%%%%

%%%%%%%%%%%%%%%%% BODY OF PAPER %%%%%%%%%%%%%%%%%%
\section{Introduction}
\noindent
It is well known that transient neutron star (NS) X-ray binaries (NSXRBs) undergo repeating outbursts driven by accretion and experience lengthy periods of quiescence when the accretion is considerably reduced. These changes in state provide a unique opportunity to probe both the properties of the compact objects and the accretion process. Several spectral states can be associated with NSXRBs including the hard, intermediate and soft state, that exhibit some overlap in spectral and timing properties with XRBs hosting black holes (BHs) (\citet{2006csxs.book...39V, 2004MNRAS.355.1105M, 2005A&A...440..207B}). Furthermore, NSXRBs are divided into two sub-classes as Z sources (with \textit{L}$_x$ $\gtrsim$ 0.5 \textit{L}$_{Edd}$) and atoll sources (with 0.01 \textit{L}$_{Edd}$ $\lesssim$ \textit{L}$_x$  $\lesssim$ 0.5 \textit{L}$_{Edd}$). The original classification, however, was based on the evolution of the softness/hardness of the source on a color-color diagram \citep{1989A&A...225..79}, where one group traces a z-shaped track as the luminosity changes, and the other group renders a c-shaped path also referred to as the 'banana' state (see also \citep{2002MNRAS.331L..47G}). The different patterns are thought to reflect a variation in the accretion rate. \\
\\
A systematic study, using a large sample of NS low-mass X-ray binaries (LMXBs) monitored by the RXTE, was performed by \citet{2014MNRAS.443.3270M} in which they showed that the NS LMXBs exhibit hysteresis-like patterns (similar to those observed in BH LMXBs \citep{1995ApJL...442....1M}) between the hard state and the soft state. The hard state is thought to be dominated by Compton scattering off energetic electrons in the corona; the soft state is associated with thermal emission from the accretion disk. They also found that the hysteresis patterns are not seen in NSXRBs at higher accretion rates where the sources remain in the thermal-dominated, low-variability state. \\   
\\
Traditionally, the states have been identified through various spectral and timing studies with a majority of them involving variability studies focusing primarily on the extraction of the fractional RMS and hardness ratios to track the transitions as the sources undergo changes in the observed count rates. For example, the hard state spectrum (in  the 2 -20 keV band) can be described by a single power law, with a photon index $\Gamma$ of $\sim 1.5$ in addition to a minor thermal contribution from the disk. The  power spectral density (PSD) usually shows significant continuum noise with fractional RMS in the range $\sim 20 -50\%$. On the other hand, the spectrum for the soft state ($\leq 2$ keV) is dominated by the disk thermal component, and the power-law component (with $\Gamma$$\sim 2-3$) is typically very weak. The fractional RMS is reduced in strength down to $\sim 5\%$. For BH transients the essence of these features is encapsulated in the hardness-intensity diagram (HID; \citet{2001ApJS...133..377, 2005A&A...440..207B}) on which the various states trace out tracks (or hysteresis-like loops) as the sources undergo transitions during outbursts. The HIDs serve as testbeds for not only the identification of the states and their transitions but are also useful in exploring the effects of accretion rate, and the evolution of the relative contributions of the thermal and power-law components to the emission process. Unfortunately, the tracks vary significantly from source to source and even within individual sources from outburst to outburst. These variations show up as local fluctuations in the intensity for any given state and thus result in considerable dispersion in HIDs as the tracks form band-like structures (similar to hysteresis loops) rather than well-defined reproducible contours.\\ 
\\
Our motive for this study is straightforward: we wish to track the spectral changes by using a temporal scale instead of a hardness ratio. The temporal parameter of interest is the minimal timescale (MTS) that was recently deployed by \citet{2020MNRAS.853.150S} in their study of GX339-4, a well studied BH transient. We focus our attention on two NSXRBs (well-studied transient atolls \citep{2002MNRAS.331L..47G}, that have undergone a number of outbursts that have excellent RXTE coverage) with the primary goal of determining whether the MTS exhibits similar behavior to that observed in GX339-4: We expect the timescale to be useful in tracking the transitions in a way that is complementary to the the traditional HIDs. We follow \citet{2020MNRAS.853.150S} and interpret the proposed timescale to imply the smallest temporal feature in the lightcurve that is consistent with a fluctuation above the Poisson noise level. In the frequency domain, this would be equivalent to the highest frequency component in the signal at or just above the noise threshold. Assuming that the (intrinsic) signal-noise threshold is different for different states, the proposed timescale implies a temporal tag for each state and associated transitions, thus providing a tool that can be used to track their dynamic evolution. Of course, a similar timescale should be accessible from PSDs; we assume this to be the cutoff frequency where the red-noise (signal) intersects the white noise (a combination of intrinsic noise associated with the source itself and an extraneous noise independent of the target, a component that is usually minimized/removed during the construction of the PSD). For cases where the broad-band (continuum) spectrum exhibits a simple 1/f${^\beta}$ behavior, a simple phenomenological model such as a powerlaw suffices to extract this cutoff frequency. However, the spectra for different spectral states can vary quite significantly in profile and complexity, thus necessitating the use of models with increasing degrees of freedom. We, on the other hand, have chosen to estimate this timescale through a wavelet decomposition (of lightcurves), where we focus on the variance of the signal; this tends to minimize the effect of the  complex features observed in some PSDs. If the MTS indeed can be taken to be the highest frequency component in the signal with the parallel assumption that the (intrinsic) signal-noise threshold is likely different for different states, then the MTS could potentially serve as a tool for tracking the evolution of the underlying continuum of those states and their transitions.\\
\\
There are several timescales that are present in accreting binary systems: typical ones of interest include the dynamical, viscous and thermal timescales. The dynamical timescale, $t_{dyn}$, is related to the Keplerian frequency $\Omega$, which is given by $\sqrt(GM/r{^3})$, where \textit{M} is the mass of the compact object, \textit{r} gives the size of the orbit, and \textit{G} is the gravitational constant. Assuming typical numbers for a given system, one 
%M = $5M_\odot$ for the BH and r varying from $5R_{g}$ to $25R_{g}$, where $R_{g}$ = $2GM/c{^2}$, 
obtains a timescale in the range $\sim$ tens to hundreds of ms. The viscous (accretion) timescale, $t_{vis}$, computed as 1/($\alpha(H/r){^2}\Omega$), assuming the standard disk with the $\alpha$ prescription for the viscosity (\citet{1973A&A....24..337S}), with typical values for $\alpha$ and \textit{H/r} (representing the vertical and radial extensions of the accretion disk) i.e., 0.1 and 0.01 respectively, yields a significantly longer time scale i.e., extending well over hundreds of seconds compared with the dynamical scale. The other important scale worth a mention is the thermal timescale, $t_{thm}$, which provides a measure of the heating and cooling rate of the disk and, as noted by \citet{2016A&A...587A..13L}, is intermediate to the other two scales, and is related to $t_{dyn}$, by a factor of 1/$\alpha$.\\ 
\\
Our targets of interest are 4U 1608-52 and Aql X-1, a pair of well studied NSXRBs. We deploy a wavelet-based technique (\citet{2013MNRAS.432..857M}) to extract a temporal scale for the targets over their respective range of state transitions that were observed during the 2002 and 2007 outbursts for 4U 1608-52 and the 1999 and 2000 outbursts for Aql X-1, and relate that timescale to a scale extracted from PSDs using the traditional Fourier-based methods. The layout of the paper is as follows: in Section 2, we describe the selection and extraction of the available RXTE data, as well as, the extraction details of the MTS and the cutoff frequency. In Section 3, we present the main results of our analysis and the comparison of the extracted time scales. We conclude by summarizing our main findings in Section 4.
\section{Observation and Analysis}
\noindent
We analyzed publicly available Rossi X-ray Timing Explorer/ Proportional counter Array (RXTE/PCA: \citep{1993A&AS...97..355B}) observations covering the 2002 and 2007 outbursts of the atoll 4U1608-52, and the 1990-2000 outbursts for Aql X-1 (Tables 1 and 2 respectively). We used the standard RXTE data analysis using the tools of HEASOFT V.6.26 to create lightcurves with a time resolution of $2^{-12}$s (i.e., $\sim244$ $\mu$s). We used PCA data modes from different channels with different time resolution including high resolution Good-Xenon or Event data modes that cover the full energy band (2 - 60 keV), otherwise we combined Single-Bit and Event data modes to cover the full energy band. The primary reason for using the full energy band was to obtain a measure of the background noise which would appear in the highest available channels. The high resolution lightcurves were created using standard screening criteria by applying source elevation of greater than (10), SAA exclusion time of (30) minutes, the pointing offset less than 0.02, and all PCUON. For each observation, we generated background model for bright objects using the FTOOL (pcabackest) in the IDL environment. This background model is used in determining the PSD normalization.\\
\\
The PSDs were created (following the procedures of \citet{2000MNRAS.318..361N}, \citet{2003A&A...407.1039P} and \citet{2005A&A...440..207B}) for segments of 16s duration and for Nyquist frequency of 2048 Hz using Powspec 1.0 (Xronos5.22) in the IDL environment. In order to check for self consistency, a number of PSDs were created for durations of 64s as well. Fast Fourier transforms were performed for each segment to extract a Leahy (\citet{1983ApJ...266..160L}) normalized PSD for each observation. The segments were averaged and rebinned geometrically to smooth the power spectrum. \\
\\
%The observational deadtime corrected Poisson noise was calculated and subtracted from the Leahy normalization, following \citet{2000MNRAS.318..361N} and \citet{2003A&A...407.1039P}. \\
Wavelet transformations have been shown to be a natural tool for multi-resolution analysis of non-stationary time-series \citep{Flandrin89, Flandrin92, Mallat89}. Indeed, a number of applications in X-ray astronomy have appeared in the literature (Scargle et al 1993, Steiman-Cameron et al 1997, Lachowicz and Czerny 2005, and Lachowicz and Done 2010). In this study, we extract the MTS using the methodology (see Appendix A) described by \citet{2013MNRAS.432..857M, 2013MNRAS.436.2907M}.\\
\\
In the following, we provide some background and note the salient features of the methodology: The light curve is represented by a complete set of basis functions. In the case of the Fourier series, the basis functions are sinusoidal functions. In the case of wavelets, we have a choice of many diverse functions, common examples being the Haar wavelet, the Morlet function, and the Daubechies functions. We use the term ‘basis’ somewhat loosely since the functions used are ‘scaled’ versions of an original function chosen for the task (the Haar wavelet in our case). One of the biggest differences in application of the Fourier series and the use of wavelets is that the basis functions typically used in wavelet analyses are localized in time i.e., have what’s known as ‘compact’ support i.e., they are non-zero over a finite interval and vanish elsewhere. For example, the Haar wavelet, a 'box'-like function, is finite for the time interval 0 < t <1 and is equal to 1 for t > 0, -1 for t > 0.5, and zero elsewhere. This is not true for the basis functions in the Fourier series i.e., those functions provide global coverage and are technically finite over all time intervals. One of the prime reasons for using wavelets is to extract simultaneously both time and frequency information for a given signal. In the case of the Fourier series one obtains only the frequency content but the location in time of the frequencies of interest is not available. While this is not an issue for signals that are stationary i.e., the frequencies are not varying in time but it is problematic when the signal is non-stationary (or weakly non-stationary). In this case one would like to extract not only the significant frequencies (that contribute to the power in the signal) but also their location in time i.e., are they in the early part of the signal or some later part. This is where wavelets are most useful. The signal of interest of course is the measured X-ray lightcurve. We know already that the emission process leading to the lightcurves most likely involves more than one mechanism because of the variations in the observed power spectra along with the occasional occurrence of quasiperiodic features. Moreover, we know that the accretion rate is unlikely to be constant over time especially since we know that the X-ray sources undergo spectral transitions involving a large variation in intensity. Under these conditions it is likely that the underlying process is at least weakly non- stationary thus justifying the wavelet approach.\\
\\
The main utility of wavelets arises due to two basic operations: scaling and translation. The primary wavelet (sometimes referred to as the mother wavelet) can be stretched and compressed in time altering the frequency content with each (scaling) operation. The larger the scale-factor the larger the spread of the wavelet i.e., the wavelet is stretched in time, which is equivalent to a wave of low-frequency. Conversely, the smaller the scale, the more compressed (in time) the wavelet, which corresponds to a high frequency wave. The freedom to adjust the scale of the wavelet enables one to search for time structures in a given lightcurve that match the scaled wavelet. All one needs to do is to scan the entire lightcurve in a systematic way, at different scales, to probe for time structures that match a given scale factor. The scanning of the lightcurve is achieved by the so-called translation factor which simply moves the location of the wavelet function from time zero by incremental time steps until the entire duration of the lightcurve has been covered. The result of each (translation) step is coded in terms of coefficients (essentially the product of the wavelet function and the lightcurve). These coefficients (in terms of the scale and translation factors) provide a measure of the power in the signal. Clearly one only obtains finite power where there is significant overlap (convolution) of the wavelet function and some structure at the right temporal scale in the lightcurve. The process of scanning the lightcurve for every scale is repeated and the appropriate coefficients extracted.\\
\\
We used the Haar wavelet to represent the observed lightcurves: It is the simplest wavelet of the Daubechies family, essentially a 'box' function (i.e., no built-in oscillatory and/or exponential features) with the fewest vanishing moments, the most compact support~\citep{Addison02}, and is constant over its interval similar to the model assumed in the Bayesian block method~\citep{Scargle13}. The variance of the resulting detail coefficients (d$_{jk}$; Appendix A) is used to construct a logscale diagram, a plot of log of variance vs. frequency (octaves). For comparative purposes, we performed a number of extractions of the MTS using the Meyer wavelet available in the Wavelet Toolbox in MATLAB; the Meyer wavelet has oscillatory structure with a decay feature and is significantly different in profile compared to the ‘box’-like structure of the Haar wavelet. We found the results to be in excellent agreement over a time scale ranging from milliseconds to seconds.\\ 
%%%%%%%%%%%
\begin{figure}
\hspace{-1.1cm}
\centering      
\includegraphics[scale=0.55, angle = 0]{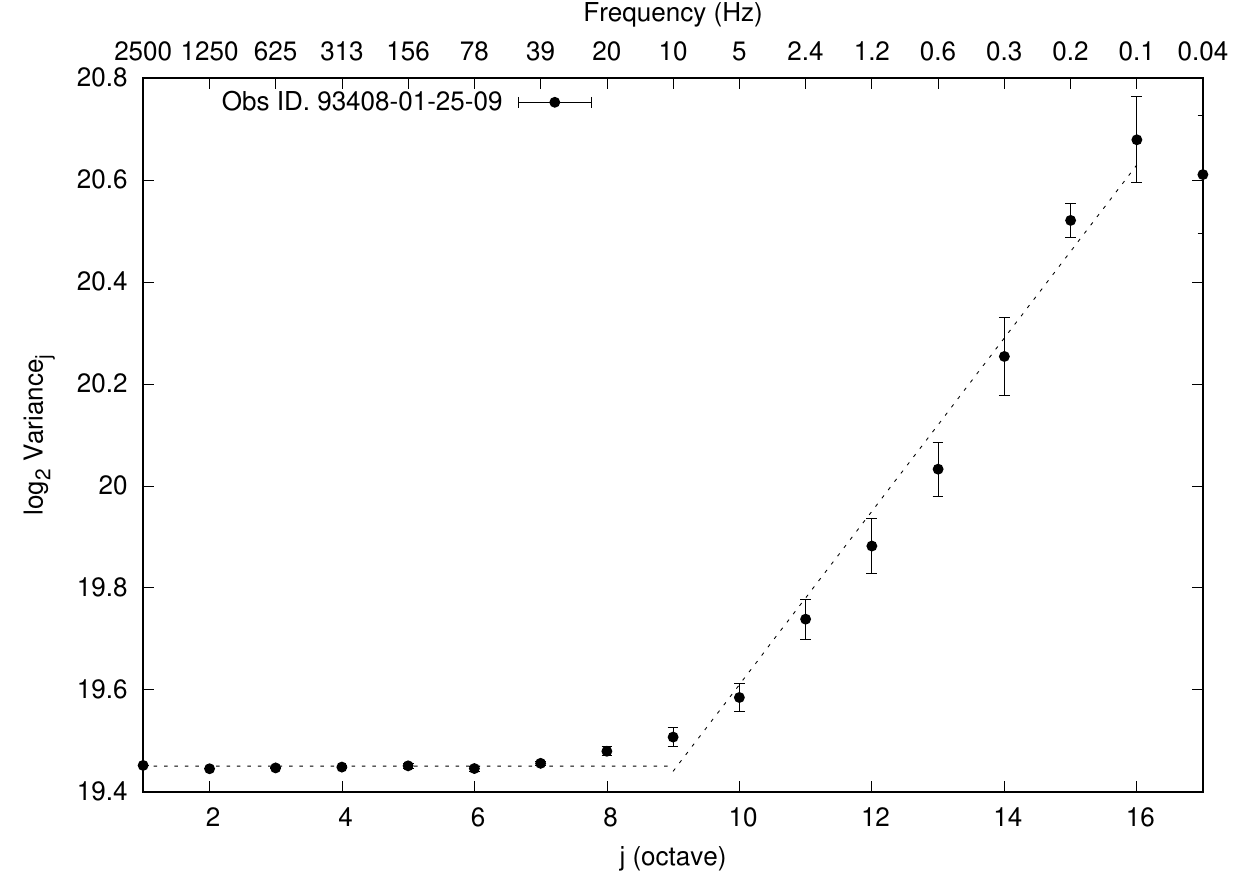}   
\caption{Logscale diagram for the observation 93408-01-25-09 (4U1608-52): variance vs octave scale. Intersection of the red-noise (sloped line) and the white-noise (flat portion) provides a measure of the MTS.}
    \label{fig1}
\end{figure}%%%%%%%%%%%%%%%%%%%%%%%
%%%%%%%%%%%
%This plot is then used to identify the regions of scaling (signal with a slope of $\lambda$) and noise (evenly distributed power as function of frequency), and extract a characteristic timescale associated with the transition between those regions.\\ 
\\
By using the logscale diagrams, we access timescales that are presumably associated with different emission processes. White-noise processes appear as flat regions while the processes generating red-noise appear as sloped regions (see Figure 1). The transition between these regions occurs at some characteristic timescale that we interpret as the minimal timescale i.e., MTS; A regression method is utilized to determine the intersection of these two regions and hence the MTS, $j$, in octaves . The octave scale is readily converted to a real time scale by using the binning time of the light curve data (i.e., the MTS (s) $\sim$ 2$^j$ x timebin). %For full details, including the procedure for subtracting the background and testing the sensitivity of the extracted result to Poisson noise, the reader is referred to the work of \citet{2013MNRAS.432..857M} and \citet{2020MNRAS.853.150S}.
\section{Results and Discussion}
\noindent
The MTS is interpreted as the smallest temporal feature in the lightcurve that is consistent with a fluctuation above Poissonian noise. Moreover, we consider this timescale to be associated more with the continuum rather than any spectral feature such as a QPO or some other resonant-like behavior in the underlying spectrum. In the frequency domain the MTS implies equivalence to the highest frequency component in the signal at or just above the noise threshold. In order to explore this implied connection, we use the standard Fourier technique to transform the lightcurves into PSDs and use a simple (broken) powerlaw (BPL) model to extract the cutoff frequency for each PSD. The cutoff frequency corresponds to the signal-noise threshold (defined by the intersection of the red and white noise components). An example of a PSD fit with the BPL is shown in Figure 2 (upper panel). Admittedly, the BPL is not an ideal model to describe every possible PSD: Indeed, considerable care has to be taken in the fitting procedure as a fair fraction of the PSDs, especially for the soft-intermediate and soft states, convey complex profiles and frequently exhibit an evolving spectral slope that progressively becomes shallower at the higher frequencies (see lower panel of Figure 2). Nonetheless, the BPL model is convenient to use and does have a limited set of parameters, which makes the model and its  parameters (in principle) considerably simpler to interpret in terms of more physically motivated models.\\
\\
The extracted lightcurves for Aql X-1 and 4U 1608-52 are displayed in Figure 3. In addition, we show in Figure 4 the corresponding hysteresis loops as function of the hardness ratio (computed as the ratio of counts in the following bands respectively: 10 - 16 keV and 6 - 10 keV): These results are in excellent agreement with those reported by \citet{2014MNRAS.443.3270M}. Specifically, the Aql X-1 lightcurves exhibit almost identical and rapid rise for both outbursts. The decays however display different profiles, with the 1999 (blue) burst falling in rate almost immediately after reaching the peak count rate whereas the 2000 (red) burst shows a much more steady decay that proceeds in two stages, a slow decay followed by a precipitous drop in rate similar to that seen in the 1999 burst. The slow/steady portion of the decay presumably coincides with a period of sustained balance between the effects of accretion and cooling. In the case of the 4U 1608-52, the rise portion, especially for the 2007 burst (green), where more observations are available, indicates somewhat episodic accretion as the lightcurve shows structure and an overall gradual increase in count rate. The decay portions of both bursts of 4U 1608-52 show a two-stage decay, a slow steady decay over significant period of time followed by a rapid drop in count rate. \\
\\
Our main results are displayed in Figures 5 and 6 where we have plotted the extracted cutoff frequencies (converted to a time scale by simple inversion) vs MTS, and the hardness vs MTS respectively for the combined data. We make several observations regarding these results: A strong positive correlation between MTS and the cutoff frequency is clearly evident. Interestingly, the individual hysteresis loops for the two sources show significant variation as function of hardness ratios, whereas the MTS-frequency correlation appears to be universal. The slope of the best-fit line (Figure 5) to the combined data is 1.01$\pm $0.01. The timescale ranges from $\sim$ 10 ms to 10 seconds, this places the range in the thermal and viscous group of time scales, which would suggest that the accretion disk is playing a significant role in determining the spectral transitions. Taken together, these results suggest that two states dominate the transitions and both are well separated in hardness and in MTS. On a closer examination of these figures (particularly Figure 6), we note that the soft state exhibits the larger MTS, and it is the hard state that corresponds to the smallest MTS. At a first glance this seems to be at odds with the findings of \citet{2020MNRAS.853.150S}, who performed a similar study of the BHXRB GX339-4. We speculate that the resolution of this apparent discrepancy resides in the fact that unlike the NSXRBs in our study, the BHXRB (GX339-4) enters a high soft state, which is not accessible to the NSXRBs binaries (\citet{2003MNRAS.342..361N}) because of the presence of the NS surface. Furthermore, we emphasize that the best-fit slope is unity with an offset that is consistent with zero (within 3$\sigma$). This provides strong evidence that the two time scales are equivalent. 
\begin{figure}
\hspace{-0.40cm}
\centering      
\includegraphics[scale=0.23, angle = 0]{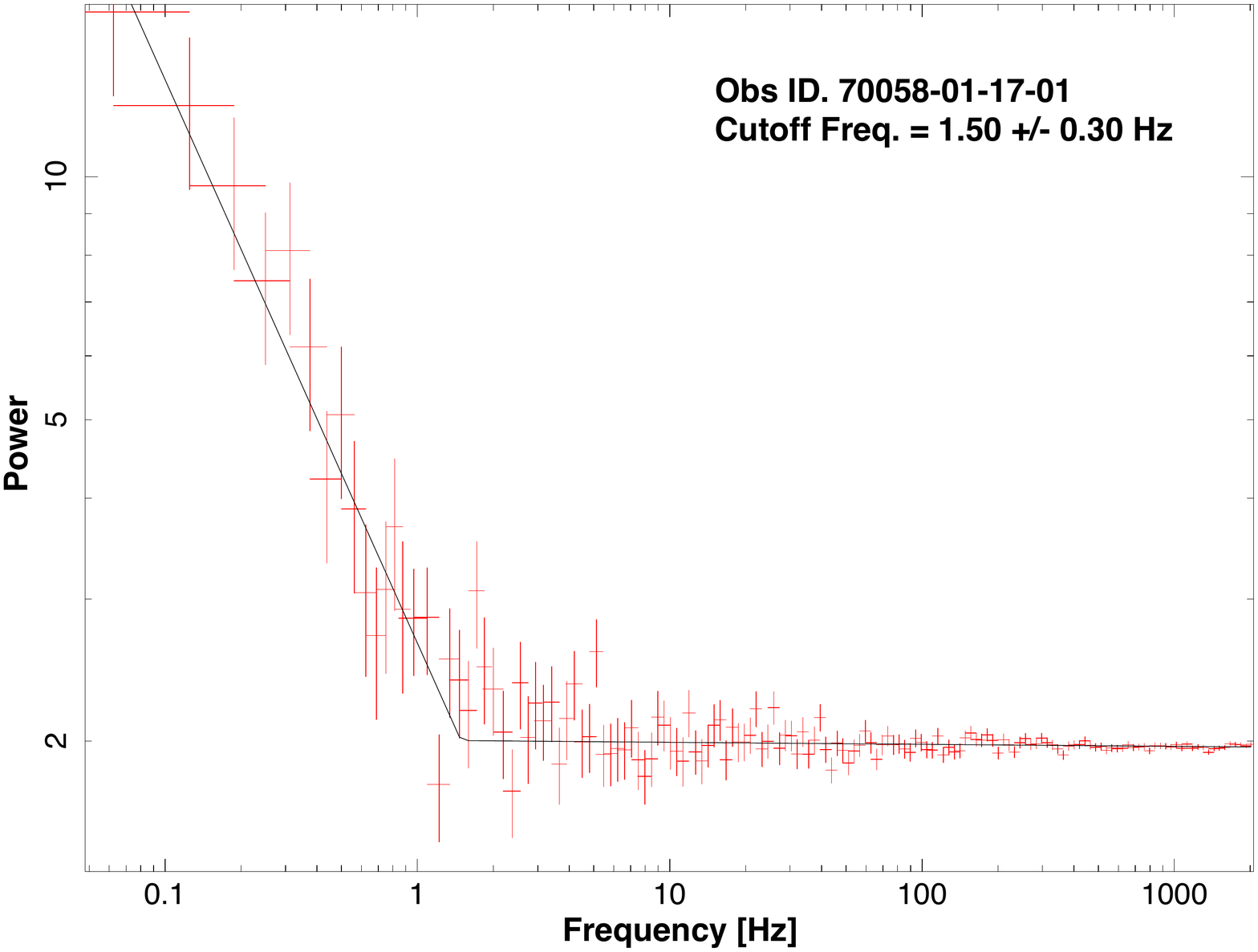}
\includegraphics[scale=0.23, angle = 0]{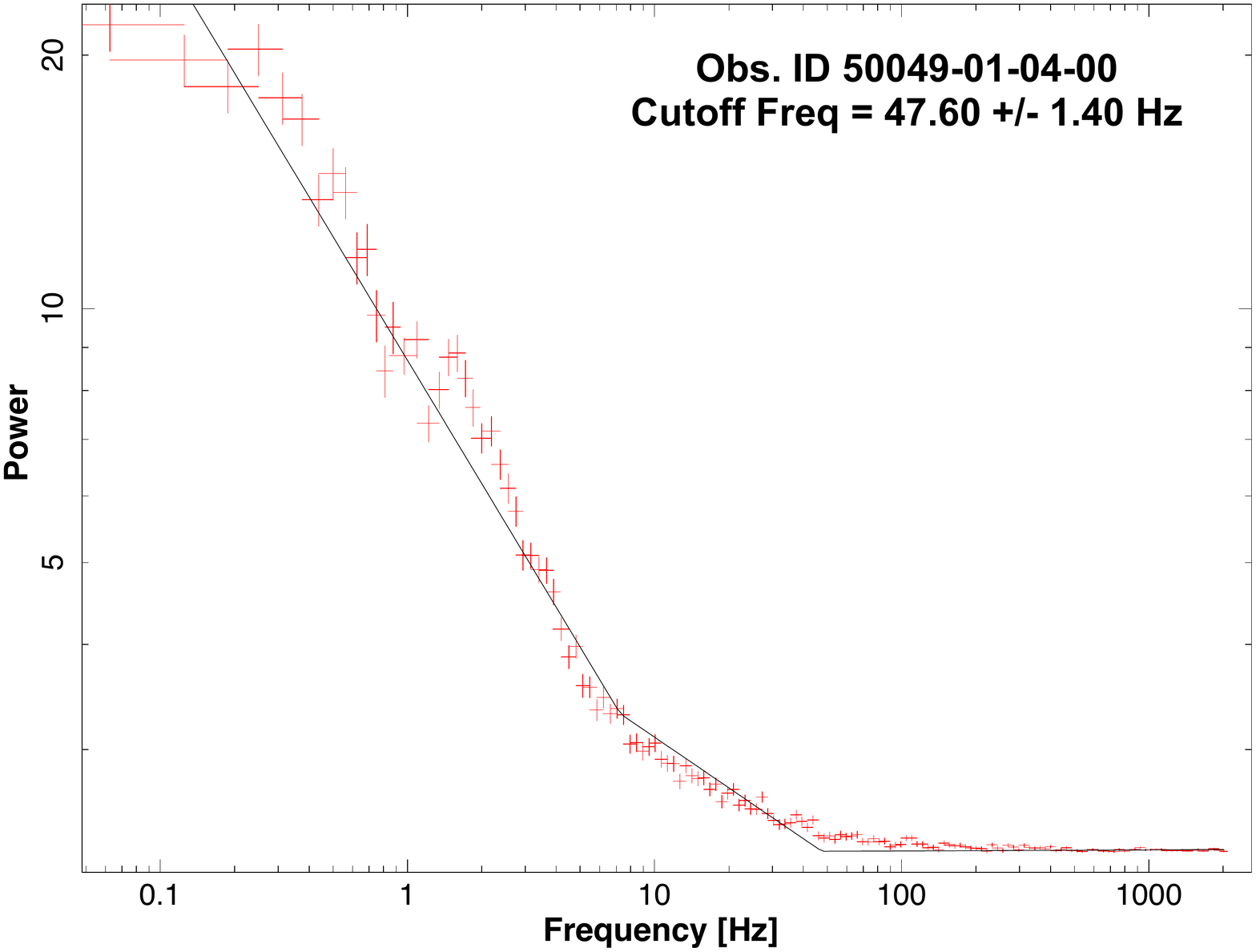}   
\caption{PSDs for the observations 70058-01-17-01 (4U1608-52; upper panel) and 50049-01-04-00 (Aql X-1; lower panel) respectively, fitted with the BPL. Notice the change of slope of the fit in the lower panel. }
\label{fig2}
\end{figure}
%%%%%%%%%%%
%%%%%%%%%%%
\begin{figure}
\hspace{-1.1cm}
\centering      
\includegraphics[scale=0.60, angle = 0]{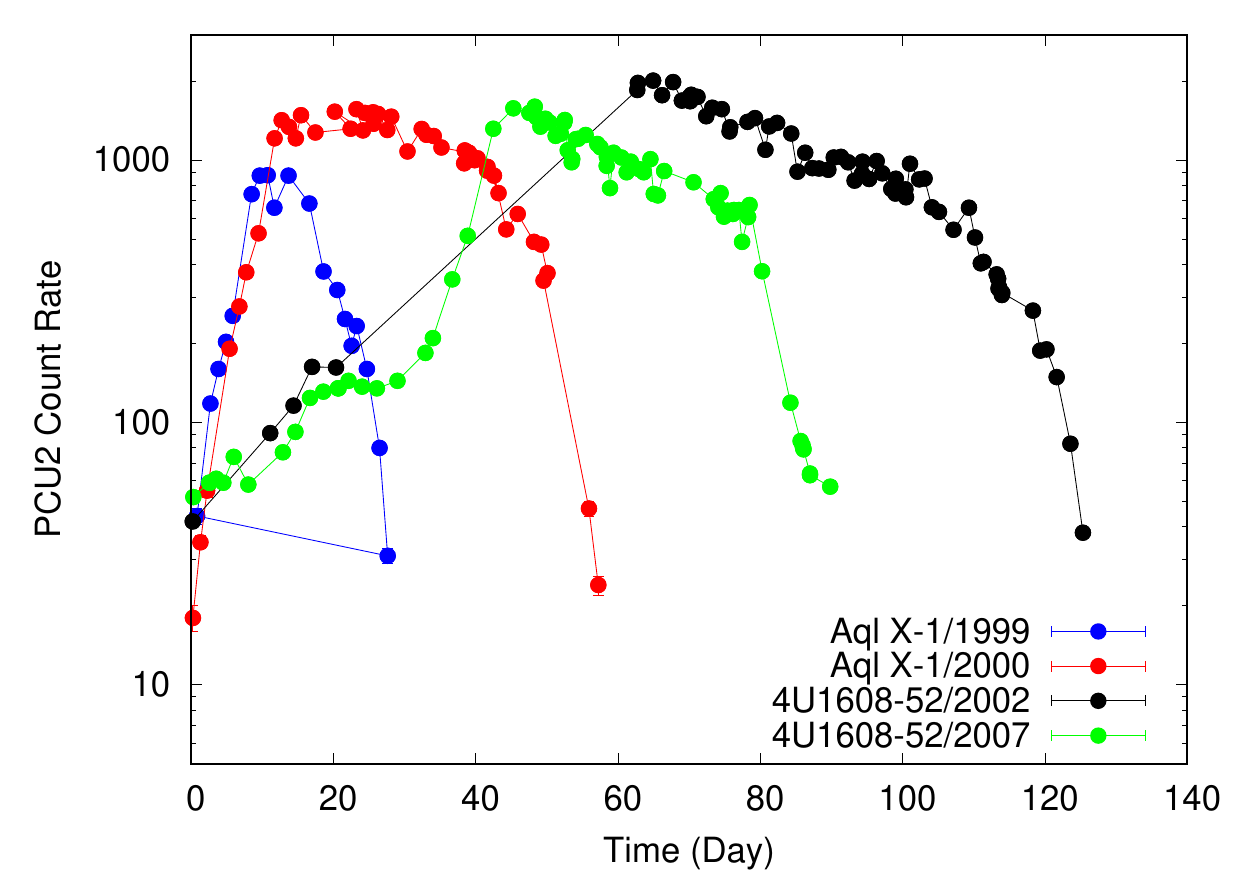}   
\caption{Lightcurves of Aql X-1 and 4U 1608-52. Each point represents the average count rate per observation (for PCU2: the energy band was 2-15 keV.)}
\label{fig3}
\end{figure}%%%%%%%%%%%%
%%%%%%%%%%%
%%%%%%%%%%%
\begin{figure}
\hspace{-1.1cm}
\centering      
\includegraphics[scale=0.60, angle = 0]{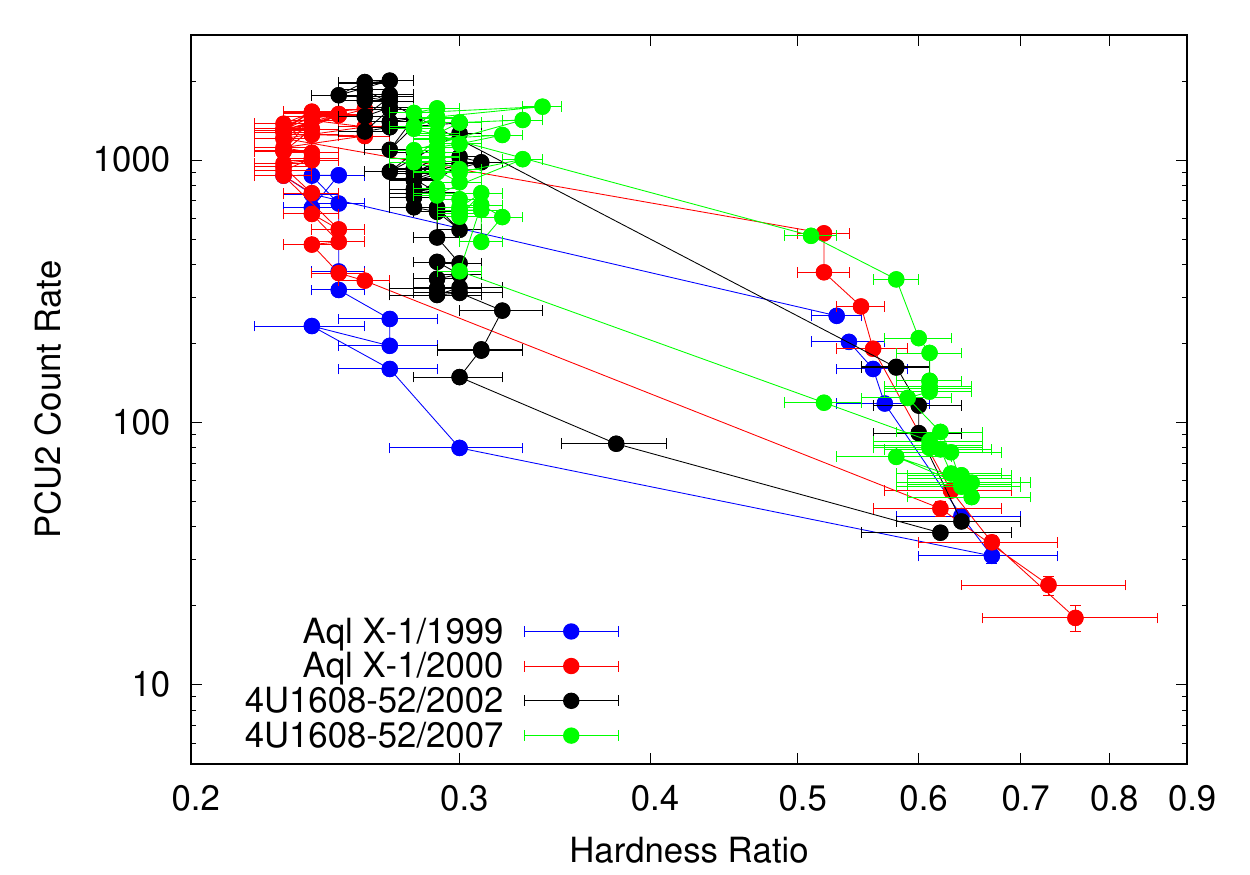}   
\caption{PCU2 count rate vs Hardness ratio for several hysteresis loops: hard and soft states are well separated.}
\label{fig4}
\end{figure}%%%%%%%%%%%%
%%%%%%%%%%%
%%%%%%%%%%%
\begin{figure}
\hspace{-1.1cm}
\centering      
\includegraphics[scale=0.33, angle = 0]{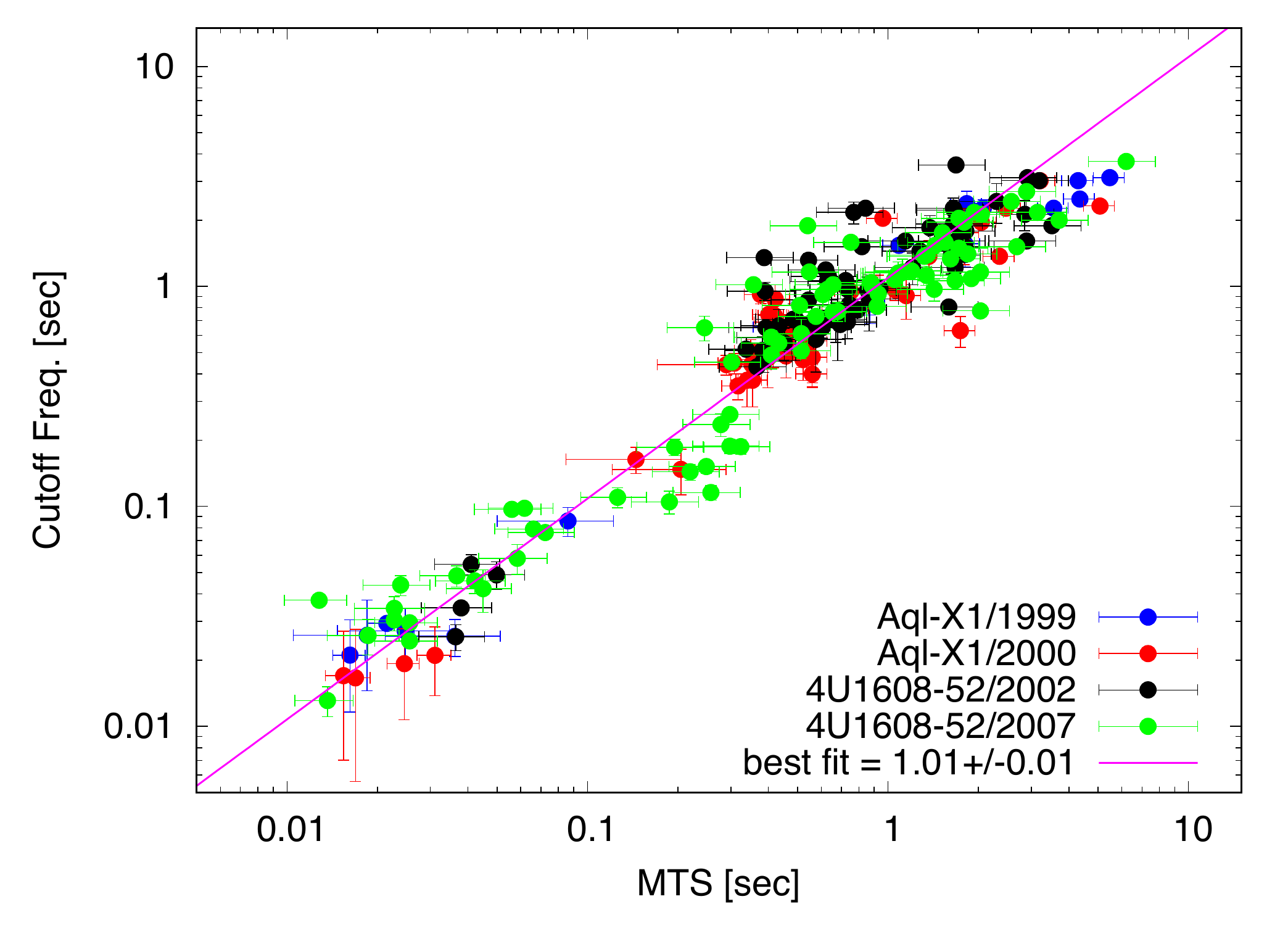}   
\caption{PSD cutoff frequencies (as a time scale) vs the corresponding MTS. The solid line is the best fit to the data.}
    \label{fig5}
\end{figure}%%%%%%%%%%%%
%%%%%%%%%%%
\begin{figure}
\hspace{-1.1cm}
\centering      
\includegraphics[scale=0.60, angle = 0]{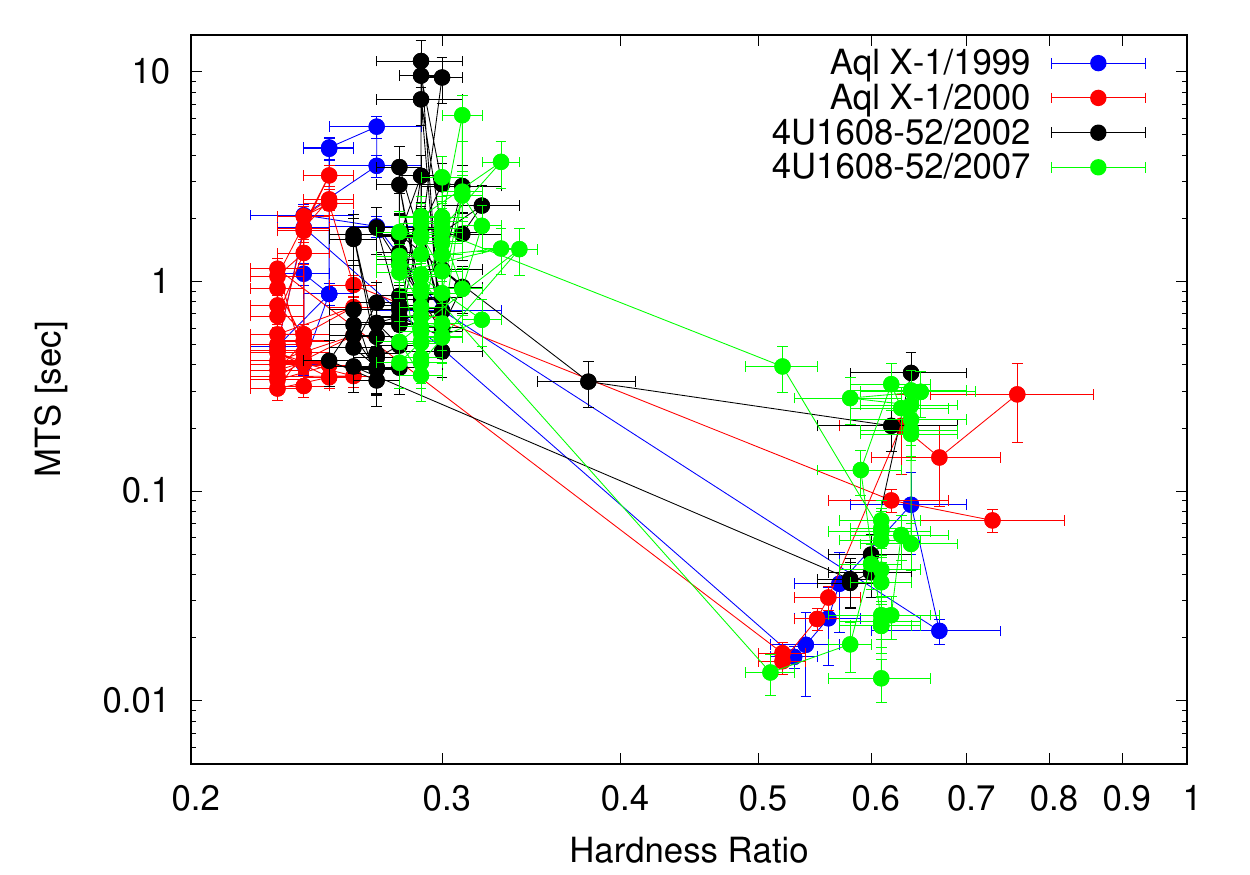}   
\caption{MTS vs Hardness ratio for several hysteresis loops: hard and soft states are well separated.}
\label{fig6}
\end{figure}%%%%%%%%%%%%
\section{Summary and Conclusions}
\noindent
In this paper, we have analyzed the available archival RXTE data for the 2002-2007 and 1999-2000 outbursts respectively for the NSXRBs 4U 1608-52 and Aql X-1.  Following the procedures described by \citet{2000MNRAS.318..361N}, \citet{2003A&A...407.1039P} and \citet{2005A&A...440..207B}, we have constructed PSDs and have extracted the cutoff frequencies by fitting the PSDs with a BPL model. We used the technique developed by \cite{2013MNRAS.432..857M}, that employs wavelet basis states, to extract a minimal time scale for lightcurves constructed for all the observations of 4U1608-52 and Aql X-1. We summarize our main findings as follows:
\begin{itemize}
\item We find a strong correlation between the MTS and the respective cutoff frequencies of the corresponding PSDs for both sources
\item Although the hysteresis loops for the sources exhibit significant variation as function of the hardness ratios, the cutoff frequency-MTS correlation appears to be 'universal' as indicated by essentially an identical best-fit to the separate datasets
\item The best-fit slope is unity and the offset is consistent with zero (within 3$\sigma$); this provides strong evidence that the two time scales are equivalent and justify the assumption that they represent the signal-noise threshold where the noise is predominantly of an intrinsic nature 
\item The extracted timescale ranges from $\sim$ 10 ms to 10 seconds; this places the range in the thermal and viscous group of time scales, indicating a strong role of the accretion disk in the observed emission.
%\item As the cutoff frequencies represent characteristic timescales associated with physically meaningful PSDs (and states represented therein) then MTS too must represent a similar physically relatable timescale
\end{itemize}

\section*{Acknowledgements}
\noindent
This research is supported by the Scientific and Technological Research Council of Turkey (TUBITAK) through project  number 117F334. In addition, KM acknowledges financial support provided by the Egyptian Cultural and Education Bureau in Washington DC, USA. The Referee's feedback was constructive and helped to clarify a number of issues.
\section*{Data availability}
The RXTE (Rossi X-ray Timing Explorer) data analyzed in the course of this study are available in HEASARC, the  NASA's Archive of Data on Energetic Phenomena. The data underlying this article are available in the article and in its online supplementary material.
%%%%%%%%%%%%%%%%%%%%%%%%%%%%%%%%%%%%%%%%%%%%%%%%%%
%%%%%%%%%%%
\begin{table}
	\centering
	\caption{Cutoff frequency and MTS for 2002 and 2007 outbursts in 4U1608-52. Full version of the table is available online.}
	\label{tab:example_table}
	\begin{tabular}{rrrrrr} % four columns, alignment for each
	\hline
Obs. Id. & Cutoff Freq. & $\delta$ Cutoff Freq.&  $\tau$ & $\delta\tau^{\pm}$   \\
	\hline
& Hz & Hz & sec & sec   \\
\hline\hline
& & {\bf 2002 outburst}  & & \\
\hline
70058-01-07-00	& 	2.32	& 	0.37	& 	0.367	& 	0.092	\\
70058-01-10-00	& 	20.47& 	1.43	& 	0.050	& 	0.012		\\
70058-01-11-00	& 	18.33& 	1.73	& 	0.041	& 	0.010	\\
70058-01-12-00	& 	39.20& 	2.93	& 	0.037	& 	0.009	\\
70058-01-13-00	& 	28.95& 	5.79	& 	0.038	& 	0.010	\\
%70059-01-01-03	& 	1.54	& 	0.37	& 	0.392	& 	0.098	\\
%70059-01-01-02	& 	1.03	& 	0.21	& 	0.732	& 	0.183	\\
%70058-01-15-00	& 	1.93	& 	0.38	& 	0.337	& 	0.084	\\
%70058-01-16-00	& 	1.71	& 	0.21	& 	0.418	& 	0.104	\\
%70058-01-17-00	& 	1.40	& 	0.57	& 	0.621		& 	0.155		\\
\hline
\end{tabular}
\end{table}
%%%%%%%%%%%

%%%%%%%%%%%%%%%%%%%%%%%%%%%%%%%%%%%%%%%%%%%%%%%%%%
%%%%%%%%%%%
\begin{table}
	\centering
	\caption{Cutoff frequency and MTS for 1999 and 2000 outbursts in Aql X-1. Full version of the table is available online.}
	\label{tab:example_table}
	\begin{tabular}{rrrrrrr} % four columns, alignment for each
	\hline
Obs. Id. & Cutoff Freq. & $\delta$ Cutoff Freq.&  $\tau$ & $\delta\tau^{\pm}$   \\
	\hline
& Hz & Hz & sec & sec   \\
\hline\hline
& & {\bf 1999 outburst}  & & \\
\hline
40047-01-01-00	& 	47.53	& 	1.06	& 	0.016	& 	0.002	\\
40047-01-01-02	& 	0.64	& 	0.03	& 	1.809	& 	0.213	\\
40047-02-01-00	& 	1.46	& 	0.06	& 	0.491	& 	0.058	\\
40047-02-02-00	& 	1.14	& 	0.05	& 	0.872	& 	0.103	\\
40047-02-03-00	& 	1.53	& 	0.13	& 	0.404	& 	0.048	\\
%40047-02-03-01	& 	0.65	& 	0.10	& 	1.085	& 	0.128	\\
%40047-02-04-00	& 	1.20	& 	0.11	& 	0.874	& 	0.103	\\
%40047-02-05-00	& 	0.33	& 	0.07	& 	4.307	& 	0.508	\\
%40047-03-01-00	& 	0.40	& 	0.08	& 	4.357	& 	0.514	\\
%40047-03-02-00	& 	0.32	& 	0.06	& 	5.468	& 	0.644	\\
\hline
\end{tabular}
\end{table}
%%%%%%%%%%%



%\bibliographystyle{mnras}

%\end{document}
%\begin{thebibliography}{99}
%\end{thebibliography}

%%%%%%%%%%%%%%%%%%%%%%%%%%%%%%%%%%%%%%%%%%%%%%%%%%

%%%%%%%%%%%%%%%%% APPENDICES %%%%%%%%%%%%%%%%%%%%%

\appendix

\section[]{Wavelet Transforms}\label{WaveletTransforms}
%Wavelet transformations have been shown to be a natural tool for multi-resolution analysis of non-stationary time-series \citep{Flandrin89,Flandrin92,Mallat89}. Wavelet analysis is similar to Fourier analysis in many respects but differs in that wavelet basis functions are well-localized, \emph{i.e.} have compact support, while Fourier basis functions are global. Compact support means that outside some finite range the amplitude of wavelet basis functions goes to zero or is otherwise negligibly small \citep{Percival00}. 
%In principle, a wavelet expansion forms a faithful representation of the original data, in that the basis set is orthonormal and complete.
%\subsubsection{Discrete Dyadic Wavelet Transforms}
%Given the discrete nature of the lightcurve data, we employ a discrete wavelet analysis. By construction, the discrete wavelet transform is a multi-resolution operation \citep{Mallat89}. Such wavelets, denoted $\psi_{j,k}(t)$, form a dyadic basis set, i.e. wavelets in the set have variable widths and variable central time positions. \\
The wavelet analysis employed in this study, as with the fast Fourier transform, begins with a light curve with $N$ elements,
\begin{equation}
X_i=\{X_0 \mathellipsis X_{N-1}\},
\end{equation}
where $N$ is an integer power of two. The light curve is convolved with a scaling function,
$\phi_{j,k}(t_i)$, and wavelet function, $\psi_{j,k}(t_i)$ which are rescaled and translated versions of the original scaling and wavelet functions $\phi(t_i)=\phi_{0,0}$, and $\psi(t_i)=\psi_{0,0}$. Translation is indexed by $k$ and rescaling is indexed by $j$. The rescaling and translation relation is given by
\begin{equation}
\psi_{j,k}(t) = 2^{-j/2}\psi(2^{-j}t-k).
%\label{eq:dyad_wavelet}
\end{equation}
The scaling function acts as a smoothing filter for the input time-series and the wavelet function probes the time-series for detail information at some time scale, $\Delta t$, which is twice that of the finest binning of the data, $T_{\rm bin}$. In the analysis, the time scale is doubled $\Delta t \rightarrow 2\Delta t$ and the transform is repeated until $\Delta t = NT_{\rm bin}.$
%%%%%
In this analysis we choose the Haar~\citep{Addison02} scaling/wavelet basis because it has the smallest possible support, has one vanishing moment, and is equivalent to the Allan variance~\citep{Howe95}, allowing for a straightforward interpretation. Convolving the light curve, $X$, with the scaling functions yields approximation coefficients,
\begin{equation}
a_{j,k} = \langle \phi_{j,k}, X \rangle.
\end{equation}
Interrogating $X$ with the wavelet basis functions yields scale and position dependent detail coefficients: The coefficients of the transform are written as
\begin{equation}
%\[
d_{j,k} = \langle \psi_{j,k}, X \rangle.
%\]
\end{equation}
The values which $j$ and $k$ assume obey the dyadic partitioning  scheme~\citep{Mallat89,Addison02,Percival00}. That is, for a time series whose number of elements is given by $N=2^m$,
%\[
\begin{equation}
0\leq j\leq m-1, and 0\leq k \leq 2^j-1.
\end{equation}
Applying the dyadic partitioning scheme removes any redundant encoding of information by the wavelet transform coefficients and guarantees orthogonality among the wavelet basis for any change in $j$ or $k$,
\begin{equation}
\langle \psi_{j,k},\psi_{j',k'}\rangle=\delta_{j,j'}\delta_{k,k'}.
\label{eq:dwt-ortho}
\end{equation}
It is interesting to point out that for the trivial $N=2$ case the Haar wavelet transform and the Fourier transform are identical.
\subsection{Logscale Diagrams and Scaling}\label{sec:LD}
Logscale diagrams are useful for identifying scaling and noise regions. Construction of a logscale diagram for each GRB proceeds from the variance of detail coefficients \citep{Flandrin92},
\begin{equation}
\rho_j = \frac{1}{n_j}\sum_{k=0}^{n_j-1}|d_{j,k}|^2,
\label{eq:waveletVariances}
\end{equation}
where the $n_j$ are the number of detail coefficients at a particular scale, $j$. A plot of $\log_2$ variances versus scale, $j$, takes the general form
\begin{equation}
\log_2 \ (\rho_j)  = \alpha j+{\rm constant},
\end{equation}
and is known as a logscale diagram. A linear regression is made of each logscale diagram and the slope parameter, $\alpha$, (depicting a measure of scaling) is estimated.  White-noise processes appear in logscale diagrams as flat regions while non-stationary processes appear as sloped regions with the following condition on the slope parameter,$\alpha>1$~\citep{Abry03,Percival00,Flandrin89}. 
\subsubsection{Statistical Uncertainties and Spurious Artifacts}
We have considered the statistical uncertainties in the light curve by a typical bootstrap approach in which the square root of the number of counts per bin is used to generate an additive poisson noise. A new poisson noise is considered for each iteration through the bootstrap process.
%\subsubsection{Circular Permutation}
%\label{circperm}
Spurious artifacts due to incidental symmetries resulting from accidental misalignment \citep{Percival00,Coifman95translation-invariantde-noising} of light curves with wavelet basis functions are minimized by circularly shifting the light curve against the basis functions. Circular shifting is a form of translation invariant de-noising~\citep{Coifman95translation-invariantde-noising}.
It is possible a shift will introduce additional artifacts by moving a different symmetry into a susceptible location. Thus, our approach is to circulate the signal through all possible values,
or at least a representative sampling, and then take an average over the cases which do not show spurious correlations.
%\subsubsection{Reverse-Tail Concatenation}
%\label{sec:rtc}
Both discrete Fourier and discrete wavelet transformations imply an overall periodicity equal to the full time-range of the input data. This can be interpreted to mean that for a series of $N$ elements,
$\{X_0,X_1\mathellipsis X_{N-1}\}$ then $X_0$ is made a surrogate for $X_N$ and $X_1$ is made a surrogate for $X_{N+1}$, and so forth. This assumption may lead to trouble if $X_0$ is much different from $X_{N-1}$.  In this case, artificially large variances may be computed. Reverse-tail concatenation minimizes this problem by making a copy of the series which is then reversed and concatenated onto the end of the original series resulting in a new series with a length twice that of the original. Instead of matching boundary conditions like,
\begin{equation}
X_0, X_1,\ldots,X_{N-1}, X_0,
\end{equation}
we match boundaries as,
\begin{equation}
X_0, X_1,\ldots X_{N-1},X_{N-1}, \ldots , X_1, X_0.
\end{equation}
Note that the series length has thus artificially been increased to $2N$ by reversing and doubling of the original series. Consequently, the wavelet variances at the largest scale in a logscale diagram reflect this redundancy. This is the reason the wavelet variances at the largest scale are excluded from least-squares fits of the scaling region.\\ 
\\
As noted, we use a bootstrap technique to estimate the uncertainties on $\rho$ (the variance of the signal) calculated from the detailed coefficients. The basic algorithm is the following: from each original lightcurve, we select a random sample of data points (equal to the dimension of the lightcurve). To this new set of data points, we add a poissonian contribution according to the count rate per each bin --- this new lightcurve is then run through our wavelet analyzer and the detailed coefficients are extracted which enables the calculation of $\rho$ for each scale factor \textit{j}. The process is repeated 100 times leading to a set of 100 $\rho$’s for each scale factor. The mean, minimum and maximum $\rho$’s are extracted for each scale factor, and finally, we compute the standard deviation for each scale factor in the standard way. \\
%%%%%%%%%%%%%%%%%%%%%%%%%%%%%%%%%%%%%%%%%%%%%%%%%%
% Don't change these lines
\bsp	% typesetting comment
\label{lastpage}
\end{document}